\documentclass[twocolumn,showpacs,preprintnumbers,amsmath,amssymb]{revtex4}

\usepackage{graphicx}
\usepackage{dcolumn}
\usepackage{bm}

\begin{document}

\preprint{APS/123-QED}

\title{Antiferromagnetic order driven by the molecular orbital order of C$_{60}$ 
in $\alpha'$--tetra--$kis$--(dimethylamino)--ethylene--C$_{60}$ }

\author{Takashi Kambe}
  \email{kambe@science.okayama-u.ac.jp}
\author{Koichi Kajiyoshi}
 \altaffiliation[Present address: ]{Rigaku Co., Tokyo, Japan}
\author{Motoyasu Fujiwara}
 \altaffiliation[Present address: ]{Institute of Molecular Science, Okazaki, Japan}
\author{Kokichi Oshima}
\affiliation{
Graduate School of Natural Science and Technology, 
Okayama University, 3--1--1 Tsushimanaka, Okayama 700--8530, Japan
}

\date{\today}

\begin{abstract}
We have studied the ground state of a fullerene--based magnet, the $\alpha'$--phase 
tetra--$kis$--(dimethylamino)--ethylene--C$_{60}$ ($\alpha'$--TDAE--C$_{60}$), 
by electron spin resonance (ESR) and magnetic torque measurements. 
Below T$_N=7$ K, non--paramagnetic field dependent resonances with a finite excitation gap (1.7 GHz) are 
observed along the $a$--axis. 
Strong enhancement in their intensity as  temperature is decreased is inconsistent with  excitation from a singlet state, 
which had been proposed for the $\alpha'$--phase ground state. 
Below T$_N$, non--quadratic field dependence of magnetic torque signal is also observed in contrast to  
quadratic field dependence in the paramagnetic phase. 
The angle--dependent torque signals below T$_N$ indicate the existence of an anisotropy of the bulk magnetization. 
From both experiments, we propose an antiferromagnetic ground state driven by the cooperative orientational ordering of 
C$_{60}$ in the $\alpha'$--TDAE--C$_{60}$. 
\end{abstract}

\maketitle

Molecular orbital ordering plays a key role in generating coherent magnetic interactions
among localized spins on a fullerene cage \cite{iwasa1} . 
The lowest--unoccupied molecular orbital (LUMO) of the C$_{60}$ cage is known to have a threefold orbital degeneracy. 
This degeneracy is partly removed by a distortion of the C$_{60}^-$ cage due to  Jahn--Teller (JT) coupling. 
The cooperative arrangement between the wave functions of JT--distorted cages should be responsible for 
the coherent intermolecular magnetic interactions. 
Namely, one expects that 
the {\it ferro(antiferro)--rotative} orbital ordering of neighboring cages leads {\it antiferro(ferro)} 
magnetic correlation between cages, respectively. 
This type of ferromagnetism in the fullerene system has been discussed by Kawamoto, and 
the quasi--degeneracy on the $t_{1u}$ orbitals is advantageous for a ferromagnetic interaction \cite{kawamoto}. 
The close coupling between the spin and orbital degrees of freedom is a conspicuous characteristic of fullerene--based magnets in comparison with other molecular magnets. 
However, it is very interesting to find a striking resemblance to the case of  orbitally--ordered inorganics \cite{kugel}. 

Tetra--kis--(dimethylamino)--ethylene (TDAE) C$_{60}$ is one of the fullerene--based magnets \cite{allemand}. 
It has two polymorphic crystal structures, the  $\alpha$ and $\alpha'$ phases. 
The $\alpha$--phase is a bulk ferromagnet with the highest transition temperature, $T_c=16$ K, 
among organic ferromagnets \cite{allemand, arcon}. 
Our structural study revealed that the ferromagnetic interaction of the $\alpha$--phase 
is grounded on  structural peculiarities \cite{kambe1,kambe2}. 
Only the $\alpha$--phase shows a structural phase transition, around $T_s=170$ K. 
Below $T_s$, the low--temperature (LT) unit cell contains two crystallographically inequivalent C$_{60}$ sites, 
whose orbitals alternately interact with each other in all nearest--neighbor directions. 
We proposed the LT {\it antiferro--rotative} structure for these $t_{1u}$ orbitals. 

On the contrary, the $\alpha'$--phase crystals show no evidence of a structural phase transition down to 25 K. 
The unit cell of the $\alpha'$--phase contains one C$_{60}$, and the LT structure allows for the {\it ferro--rotative} arrangement of $t_{1u}$ orbitals 
within the $ab$--plane and along the $c$--axis. 
The concept of {\it molecular orbital ordering} was  first introduced for the antiferromagnetic (AF) ground state of 
the ammoniated alkali C$_{60}$ salt NH$_3$K$_3$C$_{60}$ \cite{ishii,tou,margadonna}. 
It was proposed that the $t_{1u}$ orbitals alternately order along the diagonal direction within the $ab$--plane, 
together with  directional ordering of K--NH$_3$ molecular pairs, which have a dielectric moment. 
The $\alpha$--phase takes a similar aspect so that the orientational ordering works in close cooperation with 
the cubical distortion of surrounding TDAE molecules. 
Accordingly, the magnetism in the fullerene--based system tightly correlates with the structural peculiarities, especially 
the orientational freedom of C$_{60}$ cages. 

In contrast to the $\alpha$--phase ferromagnet, 
very few studies have been reported on the LT magnetic ground state of the $\alpha'$--phase. 
This may be due to the instability of $\alpha'$--phase, crystals of which gradually degenerate to the stable $\alpha$--phase  at room temperature and 
begin to show ferromagnetic behavior. 
Annealing procedures hasten the irreversible degeneration toward the $\alpha$--phase. 
We were unsuccessful in distinguishing the two polymorphs by the room--temperature structural studies, in spite of 
the distinct difference in the LT structures \cite{kambe3}. 
It was reported that the magnetic susceptibility of the $\alpha'$--phase obeys a simple Curie--Weiss law 
at high temperature  with a negative Weiss constant. 
No magnetic ordering was observed on powder sample down to 1.5 K. 
However, ESR measurements on single crystals suggested a non--magnetic singlet ground state for this phase because of 
the strong decrease of electron paramagnetic resonance (EPR) intensity \cite{arcon2}. 
Thus, the ground state of the $\alpha'$--phase is still open to debate. 
As mentioned above, LT structures for two polymorphs have remarkable differences in C$_{60}$ orientation. 
For the $\alpha'$--phase, we expect an AF long range ordering associated with the ferro--rotative alignment 
of $t_{1u}$ orbitals. 
Therefore, we stress this system is eminently suitable for study a correlation between the orbital freedom and the magnetism 
in the fullerene--based magnets. 

Our aim is to search for a long range ordered phase in the $\alpha'$--phase TDAE--C$_{60}$ 
using high--sensitive magnetic torque and low--frequency ESR experiments. 
In this Letter, we show the appearance of AF spin ordering in $\alpha'$--TDAE--C$_{60}$ and 
discuss coherent magnetic interactions among the cages in relation to the LT molecular orbital ordering structure. 
Our findings signify the convertible magnetic interactions in two polymorphic TDAE--C$_{60}$, which basically 
have  identical structures. 
We also strive to attain a unified understanding of the mechanism of magnetic interactions between C$_{60}$. 

Single crystals were obtained by the usual diffusion method \cite{kambe1}. 
We selected crystals with no twinning (via X--ray diffraction) and 
also confirmed no ferromagnetic component by magnetization measurements down to 2 K 
using a Quantum Design superconducting quantum interference device (SQUID). 
For the usual frequency region, we used a Bruker ESP300e spectrometer. 
For the low--frequency (LF) region from 1 GHz to 3 GHz, we used a homebuilt loop--gapped resonator. 
Magnetic torque of one single crystal was measured using a piezo--resistive microcantilever (Seiko instruments.Ltd.). 
The detailed experimental set--ups for torque and LF--ESR experiments are described in ref. \cite{kajiyoshi}. 

\begin{figure}
\begin{center}
\includegraphics[width= \linewidth]{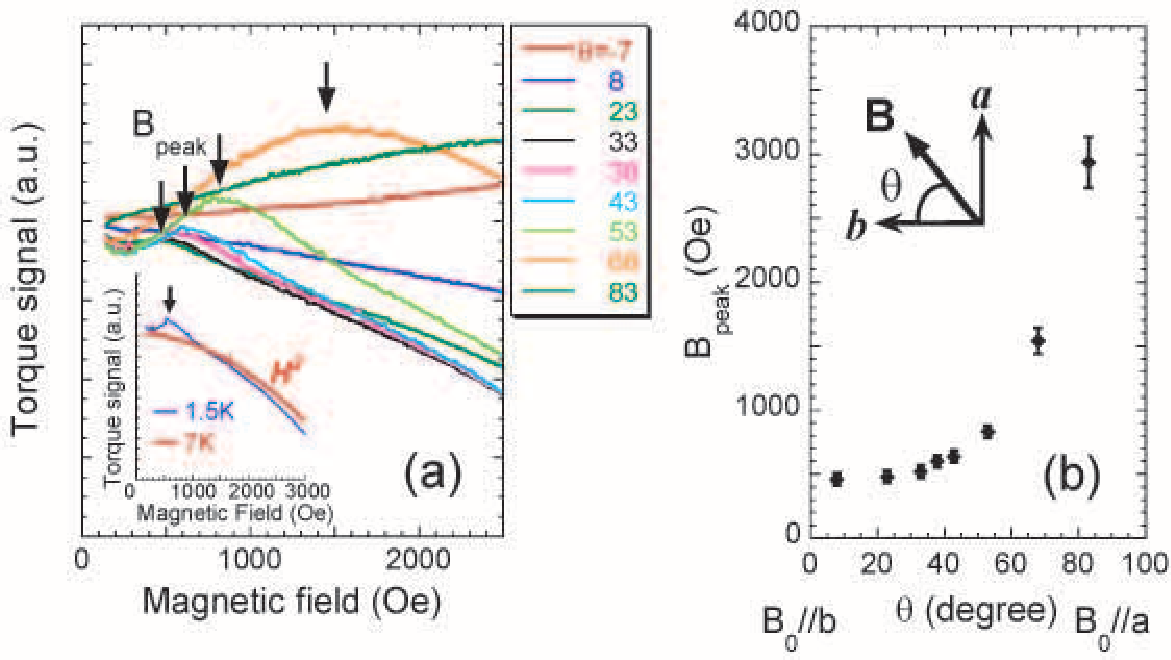}
\end{center}
\caption{
(a) Angular dependence of magnetic torque signals at $T=1.5$ K. 
The inset  shows the  torque  at 7 K (paramagnetic) and 1.5 K for $\theta=-45^{\circ}$. 
The torque curve at 7 K is proportional to the square of magnetic field strength. 
(b) Angular dependence of peak in the torque signal as a function of $\theta$. 
The inset  shows the angles between the magnetic field and the crystal axis. }
\label{torque_temp}
\end{figure}

Firstly, we present the results of magnetic torque experiments. 
The inset of figure \ref{torque_temp}(a) shows the magnetic field dependence of torque signals at 1.5 K and 7 K. 
The magnetic field is applied around $\theta=-45^{\circ}$, 
where the angles are defined in the inset of figure \ref{torque_temp}(b). 
Above 7 K, the spin system is in a paramagnetic state because the torque signal is 
proportional to the square of the magnetic field strength. 
However, below 7 K, the torque signals showed remarkable temperature dependence and had non--quadratic field dependence. 
The torque signal at 1.5 K has a clear anomaly around $H_a \sim 500 $ Oe. 
Previous ESR measurements proposed a non--magnetic ground state because of the disappearance of EPR signals. 
If the ground state is a spin--singlet, the torque signals should exactly vanish below a critical magnetic field, 
which corresponds to the spin--excitation gap. 
The observed large field dependence of torque signals in all field regions clearly deny the singlet ground state 
for the $\alpha'$--phase. 
Figure \ref{torque_temp}(a) shows the angular dependence of torque signals at $T=1.5$ K for $H \parallel ab$--plane. 
The magnetic torque signals depend on the applied magnetic field direction. 
The angular dependence of peaks in the torque signals are shown in the Fig. \ref{torque_temp}(b). 
The peak shifts toward the high field side when the magnetic field leans toward the $a$--axis. 
The angle--dependent torque signals indicate the existence of anisotropy of the bulk magnetization 
and, around $H_a$, a spin re--arrangement under the magnetic field, such as a spin--flop transition, takes place. 
The angular dependence of $H_a$ indicates a spin--hard direction along the $a$--axis. 

\begin{figure}
\begin{center}
\includegraphics[width= 0.8 \linewidth]{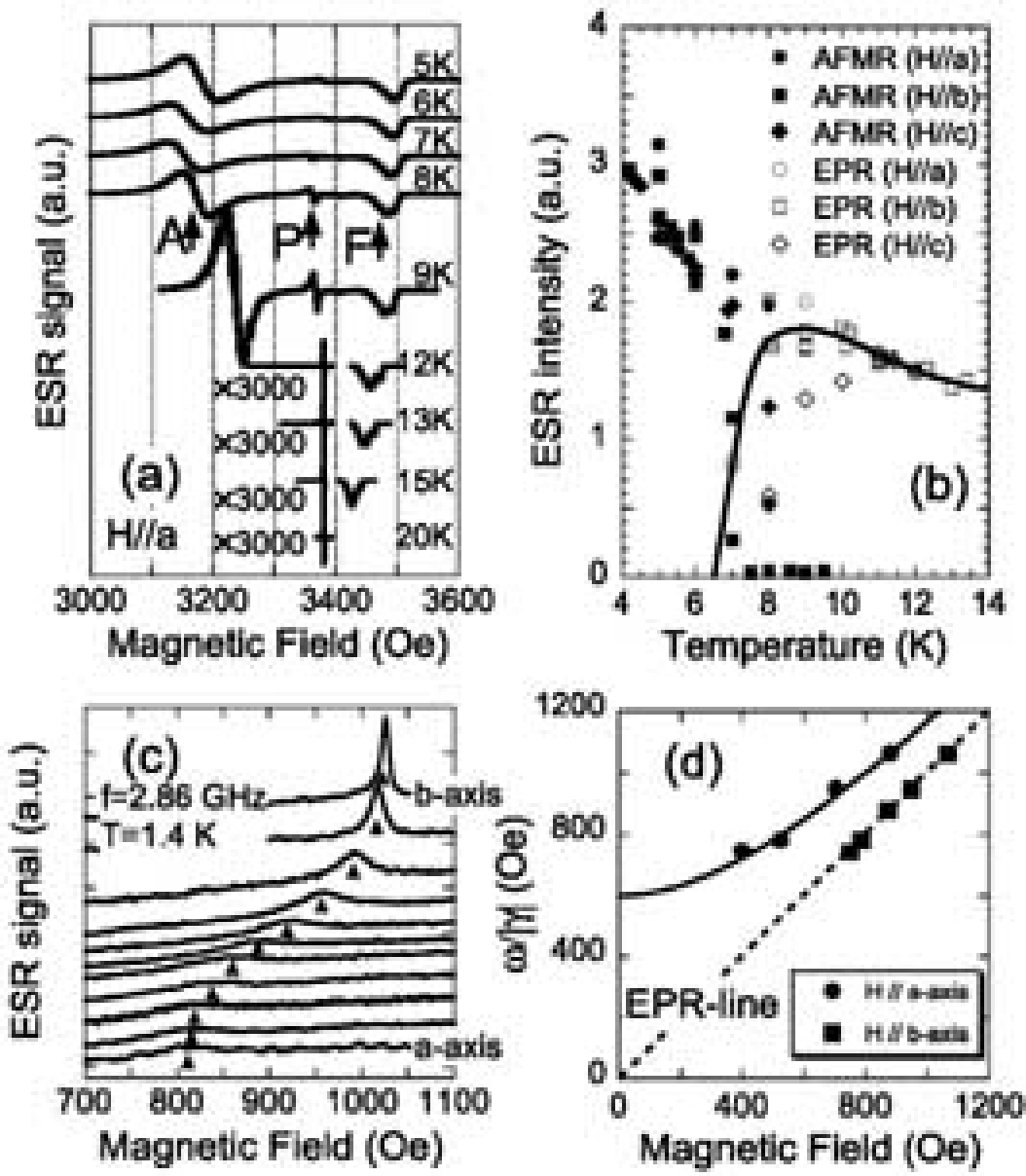}
\end{center}
\caption{
(a) Temperature dependence of ESR spectra for $H \parallel a$ at 9 GHz region. 
Three peaks ($P$, $A$ and $F$) are observed, which correspond to the EPR, AFMR and (extrinsic) FMR signals, respectively. 
(b) Temperature dependence of EPR and AFMR intensity for three crystallographic axes. 
(c) Angular dependence of AFMR signals within the $ab$--plane at 2.86 GHz.  
Each spectrum was taken every 10 degrees. 
(d) Frequency ($\omega/|\gamma|$)--field relation at T=1.5 K. 
Solid line is the calculation with the easy--plane anisotropy along the $a$--axis. 
}
\label{esr}
\end{figure}

Next, we show ESR results. 
Figure \ref{esr}(a) shows the temperature dependence of ESR spectra at the usual X--band frequency for $H \parallel a$--axis. 
Three well--separated peaks were observed in different temperature windows. 
$P$ peaks correspond to the EPR signals with  linewidths of about 20 Oe. 
Their intensities obey a simple Curie law down to 9 K. 
Below 9 K, however, the intensity rapidly decreased with decreasing temperature (see figure \ref{esr}(b)) and 
the EPR signals completely disappeared below 6 K. 
This result is consistent with the previous ESR measurements \cite{arcon2}. 
Contrary to this agreement, however, we found new absorption signals ($A$), following the disappearance of the EPR signals. 
Note that the resonance field of $A$ peaks are different from the corresponding EPR position. 
Their resonance fields markedly depend  not only on the magnetic field direction but also on the temperature. 
In figure \ref{esr}(b), the temperature dependence of the absorption intensity for the $A$ signals is also shown. 
The $A$--intensity at 9 K is about seventy times stronger than the EPR intensity. 
Strong enhancement in  intensity as  temperature decreases is inconsistent with a singlet ground state 
because the excitation from the singlet state is normally forbidden by the ESR selection--rules. 
Notice that  non--quadratic torque curves are also observed below 7 K. 
As mentioned later, we attribute the $A$ signals to antiferromagnetic resonance (AFMR) from 
the frequency--field relation of the $A$ excitation. 
Both magnetic torque and ESR experiments indicate that 
the ground state of the $\alpha'$--phase is not non--magnetic, but rather magnetic with a finite magnetic ordering temperature. 
Contrary to these intrinsic signals, $F$ peaks, which have large angular dependence, were observed below 16 K. 
This temperature corresponds to the onset temperature of ferromagnetic ordering in the $\alpha$--phase. 
The $F$ intensity increased with decreasing temperature, but was two order of magnitude weaker than that of the EPR as well as the $A$ signals. 
Thus, we assign  the $F$ peaks as originating from a degeneration of the $\alpha'$--phase on part of crystal, 
such as the surface. 
We emphasize that this crystal shows no trace of ferromagnetic component using the SQUID measurements. 

Figure \ref{esr} (c) and (d) show the angular dependence of the AFMR signals in the  low frequency region 
and the frequency--field ($\omega/\gamma$--H) diagram at 1.5 K, respectively. 
Only the resonance fields of AFMR are plotted. 
The non--paramagnetic field dependence of excitations along the $a$--axis indicates the existence of finite 
excitation energy in the spin--wave dispersion. 
The LF--ESR is profitable to observe such low--energy excitation. 
As mentioned previously, because of the strong enhancement in their intensities at lower temperature, 
these excitations can not be identified as  excitations from the singlet ground state to the triplet states. 
Accordingly, the frequency dependence of the excitation confirms the AF ordering below T$_N$ for $\alpha'$--TDAE--C$_{60}$. 
Large zero--field excitation energy exists along the $a$--axis, while there is very weak anisotropy on the 
resonance field in the $bc$--plane. 
In order to investigate these resonance field relations, we neglect the in--plane anisotropy 
within the $bc$--plane, but introduce the AF configuration with  out--of--plane anisotropy along the $a$--axis. 
The angle--dependent torque experiments support this anisotropy. 
The solid line in  figure \ref{esr} (c) is a calculation with the easy--plane anisotropy
: $\omega_2/\gamma = \sqrt{2H_A H_E}= 612$ Oe, 
where $H_A$ and $H_E$ are the out--of--plane anisotropy and the exchange molecular field, respectively \cite{nagamiya}. 
This is the first experimental evidence of a spin--wave excitation in the AF fullerene magnets. 

\begin{figure}
\begin{center}
\includegraphics[width=0.7 \linewidth]{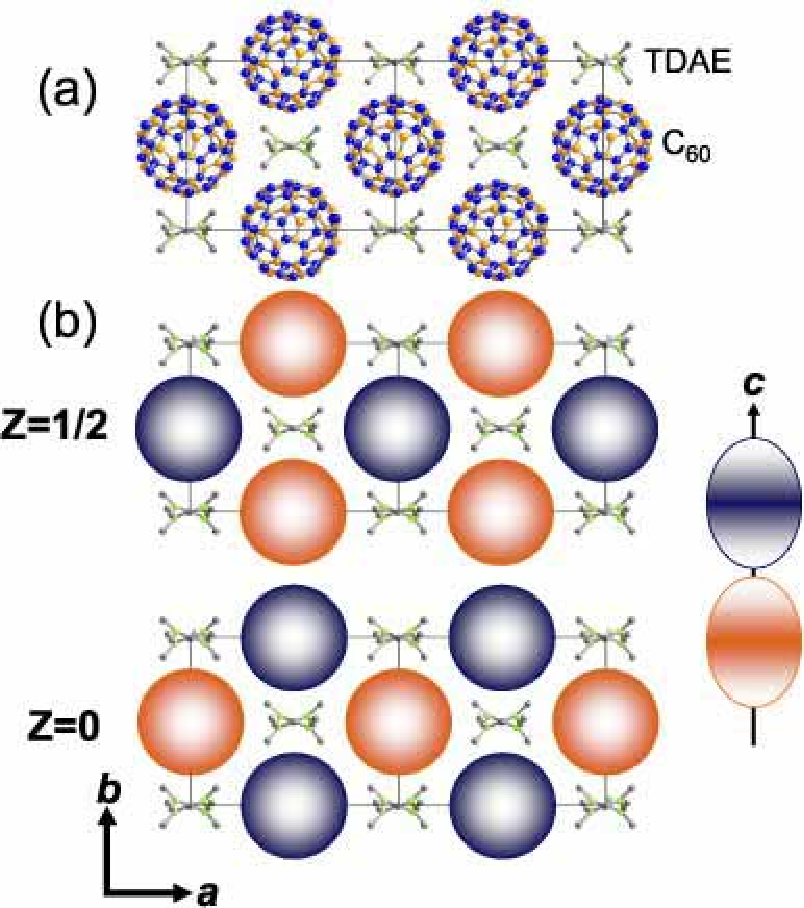}
\end{center}
\caption{(a) Low--temperature structure and (b) possible antiferromagnetic (AF) structure with molecular orbital order.
Balls represent one of the C$_{60}$ $t_{1u}$ orbitals. 
Blue and orange balls correspond to opposite spins. 
}
\label{orbital}
\end{figure}

The AF interactions among the spins on the cage should be controlled by the molecular orbital configuration and 
must dominate the dispersion relation of the magnetic excitation. 
We propose the AF spin structure driven by the cooperative ordering of $t_{1u}$ orbitals. 
Figure \ref{orbital} shows the LT structure of the $\alpha'$--phase, which includes 
one crystallographically independent C$_{60}$ in the unit cell. 
This LT structure allows ferro--rotative ordering of the $t_{1u}$ orbital due to its symmetry. 
Thus, the elongated axes of the distorted cages should align parallel to each other (see the balls in the right panel, which represent one of the $t_{1u}$ orbitals of the cage). 
This panel shows a possible molecular orbital configuration, where 
the elongated axis of the cages aligns parallel to the $c$--axis. 
This configuration is completely different from that of the $\alpha$--phase. 
The LT symmetry of the $\alpha$--phase permits an alternate configuration of $t_{1u}$ orbitals along all neighboring directions 
(Fig. 5 in Ref. \cite{kambe2}), which should be responsible for the ferromagnetic interactions. 

It should be necessary to introduce  orientational disorder 
if the symmetry of the cage is lower than that of the lattice site. 
The static orientational ordering of C$_{60}$ takes place in the LT structures of both phases. 
Thus, we expect that the distortion of the cage fits  the local symmetry of the lattice site. 
Only inversion symmetry remains at the cage site in the LT structure of $\alpha'$--phase. 
Recent molecular orbital calculations suggested that the C$_i$ distorted cage is slightly stable, but 
neither C$_i$ and C$_{2h}$ distortion can be determined \cite{Kodama,Long}. 
In fact, in [Ph$_4$As]$_2$ClC$_{60}$, the anisotropy of the ESR g--factor and  vibrational analysis 
suggest the stability of C$_{2h}$ or C$_i$ distortion \cite{bietsch,Long}. 
In the $\alpha'$--phase, on the contrary, angular variations of the ESR g--factor indicate that, below 150K, 
an axial g--factor changes to three different components 
(g$_c \sim 2.0012$, g$_b \sim 2.0003$ and g$_a \sim 2.0006$) \cite{fujiwara}. 
This anisotropy is three times lower than that of [Ph$_4$As]$_2$ClC$_{60}$. 
The largest g--factor was obtained along the $c$--axis, which corresponds to the vertical axis of the C=C double bond between 
the 6--membered rings of the cage. 
On the contrary, the elongated axis for the C$_i$ distorted cage should align with 
the axis through the five--membered rings on the cage \cite{Long, Kodama}. 
For the  $\alpha'$--phase, this direction may correspond to the $b$--axis for all cages \cite{kambe2}. 
Although the distortion (elongation) of the cage has not been experimentally determined in this TDAE--C$_{60}$ system (not only 
the $\alpha$--phase but also $\alpha'$--phase) by X--ray measurements, 
in either case of C$_i$ or C$_{2h}$, on--site symmetry possibly requires the most stable structure of the cage and 
the elongated axes of cages align parallel to each other. 
Accordingly, {\it ferro--rotative} molecular orbital ordering plays a key role to establish 
the antiferromagnetic interactions between the cages in the $\alpha'$--TDAE--C$_{60}$. 
Possible AF spin structure under this molecular orbital ordering is shown in  fig. \ref{orbital}(b), 
where the opposite spins are represented by  ball colors. 
The spin--hard direction is parallel to the $a$--axis, which is perpendicular to the elongated axes of the cages. 
This magnetic anisotropy should be mainly dominated by a dipole--dipole interaction between the spins. 
In the LT structure, the nearest--neighbor distance between the cages is along the $c$--axis, 
and the next--nearest neighbor distance is along the $b$--axis. 
The close packing configuration along the $c$--axis may yield a vertical anisotropy against the $c$--axis due to  
dipole--dipole interactions. 
This may be consistent with the AF structure with the spin--hard anisotropy along the $a$--axis. 

In summary, we have measured the magnetic torque and LF--ESR on the $\alpha'$--phase of TDAE--C$_{60}$ 
in order to search for a magnetic ground state. 
>From both experiments, we conclude the  ground state of  $\alpha'$--TDAE--C$_{60}$ is AF with $T_N\sim 7$ K and 
also propose the AF ordering structure associated with the {\it ferro--rotative} orbital ordering 
of the JT--distorted C$_{60}$, which agrees well with the molecular ordering scenario. 
Our results insist that the JT--distorted C$_{60}$ functions as the magnetic constituent element at low--temperatures, 
though the distortion of C$_{60}$ has not been verified by  LT X--ray measurements \cite{kambe3}.  The
TDAE--C$_{60}$ system provides a good example to obtain a unified understanding of the mechanism of magnetic interactions 
between C$_{60}$.
We emphasize that this system shows  convertible magnetic interactions due to 
the relative molecular orbital ordering of the JT--distorted C$_{60}$. 

This research is supported by a Grant--in--Aid for Scientific Research from the Japanese Ministry of Education, Culture, 
Sports, Science and Technology. 


%
%

\begin{references}
\bibitem{kambe3} We found a clear difference on dynamical response between the $\alpha$-- and 
$\alpha'$--phases at high temperatures. This will be appeared in a separated manuscript.
\end{references}
\end{document}